\documentclass[12pt,labels]{iopart}
\usepackage[usenames,dvipsnames]{color}
\usepackage{graphicx}
\usepackage[utf8]{inputenc}
\usepackage[T1]{fontenc}

\begin{document}
\title{Two coupled Josephson junctions: dc voltage controlled  by biharmonic current }
\author{L. Machura, J. Spiechowicz, M. Kostur, and J. Łuczka}
\address{Institute of Physics, University of Silesia, Katowice, Poland}
\ead{Jerzy.Luczka@us.edu.pl}

\begin{abstract} 
We study transport properties of two Josephson junctions coupled by an external shunt resistance. One of the junction (say, the first) is  driven by an unbiased ac current consisting of two harmonics.  The device  can rectify the ac current 
 yielding a dc voltage across the first junction. For some values of coupling strength, controlled by an external shunt resistance, a dc voltage across the second junction can be generated.  By variation of system parameters like the  relative phase or frequency of two harmonics, one can conveniently manipulate both voltages with high efficiency, e.g., changing the dc voltages across the first and second junctions 
  from  positive to negative values and {\it vice versa}. 
\end{abstract}


\maketitle

\section{Introduction}\label{sec:intro}

Transport phenomena in periodic structures by harvesting the unbiased external time--periodic
stimuli and thermal equilibrium fluctuations have been one of the hottest 
topic in nowadays science. Examples range from biology and biophysics \cite{noiseinbio}
to explain  directed motion of biological motors \cite{biomotors} and
particle transport in ion channels \cite{ionchannels},  or 
to design new separation techniques \cite{sep}, 
to meso-- and nano--physics covering newest and up-to-date experiments
with optical lattices \cite{Renzoni2005}, persistent currents in quantum rings \cite{mesorings}
and Josephson junctions \cite{jj}, to mention only a few. 
The physics of the latter has been studied for almost five decades
now \cite{Jos1964}. Devices  with Josephson junctions constitute a  paradigm of 
nonlinear systems exhibiting many  interesting 
phenomena in classical and quantum regimes.
Yet new and interesting phenomena arise as researchers are able
to use an advantage of powerful computer simulations followed by real experiments on 
junctions proving theoretically predicted findings \cite{MacLuc2007,ANM,NagSpe2008}. 
In this paper we study a system of two coupled  Josephson junctions   
driven by an external ac current having two harmonics. Such driving has been considered as a 
source of energy pumped into the system and as an agent transferring the system to non-equilibrium states. 
Transport phenomena in  non-equilibrium states are of prominent interest from both theoretical 
and experimental point of view. For example, transport processes driven by biharmonic driving have 
been intensively studied in  variety of systems and in various regimes. We can quote: 
Hamiltonian systems \cite{flach}, systems in the overdamped regime \cite{Borromeo2005a,flachepl} and  for moderate damping 
 \cite{Breymayer1984,MacLuc2010}, 
dissipative quantum systems \cite{igor}, electronic quantum pumps \cite{arachea} and quantum control of exciton states \cite{control}, 
 cold atoms in the optical lattices
 \cite{Renzoni2005,Renzoni2008,Denisov2010}, soliton physics \cite{soliton} and  
  driven Josephson junctions \cite{Monaco1990}.  
Here we intend to consider another aspect of transport in the system which consists of two coupled elements (subsystems).
In particular, we take into consideration two coupled Josephson junctions. In a more general context it is a system which can be described by 
two degrees of freedom (it can consist of two interacting particles or one particle moving on  a two-dimensional 
substrate) \cite{2D}.  The question is: What new properties of transport  can be induced by interaction between two 
elements  and under what conditions can transport be generated in the second element if the driving is applied only to the first element. 

The layout of the  paper is  as follows: In section 2, we detail the system of two  Josephson junction 
coupled by an external shunt resistance. Next, in section 3, we specify  external driving applied to 
one of the junctions. In section 4, we analyze generic properties of dc voltages across the first and second junctions.  
In section 5, we elaborate on the voltage control. Section 6  provides summary and  conclusions.

\section{Model and its dynamics}\label{sec:dynamics}

Let us consider a system of two coupled resistively shunted Josephson junctions. The equivalent circuit 
representation of the system is illustrated in figure \ref{Kir1}. The junctions  
are characterized by the critical currents $(I_{c1}, I_{c2})$,  
resistances $(R_1, R_2)$ and phases $(\phi_1, \phi_2)$ (more precisely, the phase differences 
of the Cooper pair wave functions across the junctions). The junctions are externally shunted 
by the resistance $R_3$ and driven by  external currents $I_1(t)$ and $I_2(t)$.  
We also include into our considerations  Johnson thermal noise sources $\xi_1(t), \xi_2(t)$ 
and $\xi_3(t)$ associated with the corresponding resistances $R_1, R_2$ and $R_3$.   

From the Kirchhoff current and voltage laws, and two Josephson relations one can obtain the following equations 
\begin{eqnarray}
  \label{Kir1}
  \frac{R}{R_1} \frac{\hbar}{2e}  \dot{\phi}_1 &=& (R_2+R_3) [I_1(t)- I_{c1} \sin \phi_1] 
  +  R_2 [I_2(t)- I_{c2} \sin \phi_2] \nonumber\\
  &-&(R_2+R_3)\, \xi_1(t) -R_2 \,\xi_2(t) -R_3 \,\xi_3(t),  \nonumber\\
     \frac{R}{R_2} \frac{\hbar}{2e}  \dot{\phi}_2 &=& (R_1+R_3) [I_2(t)- I_{c2} \sin \phi_2] 
  +  R_1 [I_1(t)- I_{c1} \sin \phi_1] \nonumber\\
  &-&(R_1+R_3)\, \xi_2(t) -R_1 \,\xi_1(t) +R_3 \,\xi_3(t),  
\end{eqnarray}
where the dot denotes a derivative of $\phi_i=\phi_i(t) \, (i=1, 2)$ with respect to time $t$ and $R=R_1+R_2+R_3$. We assume that
all resistors are  at the same temperature $T$  and the thermal equilibrium noise sources  are represented  by zero-mean Gaussian white noises $\xi_i(t)
\, (i =1, 2, 3)$   which are delta-correlated, i.e.  $\langle \xi_i(t) \xi_j(s)
\rangle \propto \delta_{ij} \delta(t-s)$  for $i, j \in \{1, 2, 3\}$. 
Note a symmetry property of equations (\ref{Kir1}) with respect to the change $R_1 \leftrightarrow  R_2$. 
%
\begin{figure}[t]
	\centering
	\includegraphics[width=0.6\textwidth]{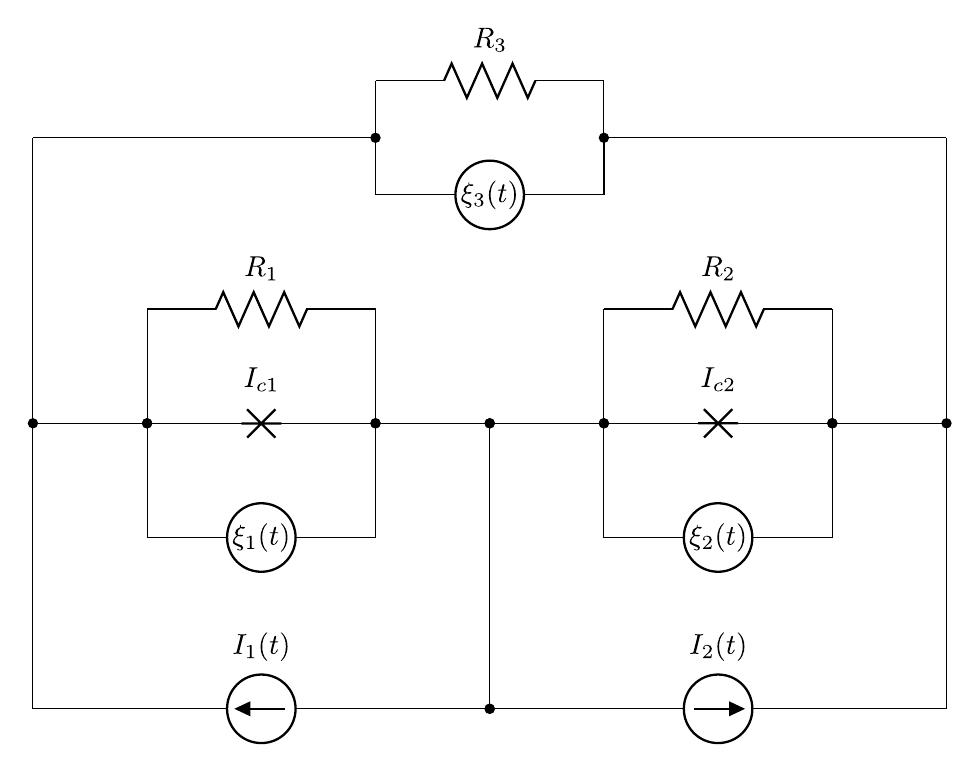}
	\caption{The system of two resistively shunted  Josephson junctions
	coupled  by  an  external  shunt resistance $R_3$ and driven by the
	currents $I_1(t)$ and
	$I_2(t)$.}
	\label{fig1}
\end{figure}
%

The limitations of the model (\ref{Kir1}) and its range of
validity are discussed e.g. in  Ref. \cite{kautz}.  In particular, we
work within the  semi-classical regime of   small junctions for which a spatial dependence of characteristics can be neglected, a displacement current accompanied with the
junction capacitance is sufficiently small and  can be ignored,  and 
photon-assisted tunneling phenomena do not contribute. 

 The dimensionless variables and parameters  can be introduced in various way. Here, we follow 
Ref. \cite{Ner21} and define the dimensionless time $\tau$ as: 
\begin{eqnarray}
  \label{time}
	\tau = \frac{2eV_0}{\hbar} t, 
\end{eqnarray}
where 
\begin{eqnarray}
  \label{voltage}
	V_0 =  I_c \frac{R_1(R_2+R_3)}{R_1+R_2+R_3}, \quad I_c=\frac{I_{c1}+I_{c2}}{2} 
\end{eqnarray}
are the the characteristic voltage and averaged critical current, respectively. 
The corresponding dimensionless form of equations (\ref{Kir1}) reads
\begin{eqnarray}
  \label{fi}
    \dot{\phi}_1 = I_1(\tau)- I_{c1} \sin \phi_1 
      +\alpha [I_2(\tau)  -  I_{c2} \sin \phi_2] + \sqrt{D}\; \eta_1(\tau),  \nonumber\\
    \dot{\phi}_2 =  \alpha \beta[I_2(\tau) -  I_{c2} \sin \phi_2]
    	       + \alpha [ I_1(\tau) - I_{c1} \sin \phi_1] + \sqrt{\alpha \beta D} \;\eta_2(\tau),    
\end{eqnarray}
where now the dot denotes a derivative with respect to dimensionless time $\tau$ and all dimensionless currents are in units of $I_c$. E.g., $I_{c1} \to I_{c1}/I_c$.  The parameter 
\begin{eqnarray}
  \label{alfa}
  \alpha = \frac{R_2}{R_2 + R_3}  \in [0, 1]
\end{eqnarray}
 is the coupling constant  between two 
junctions and can be controlled by the external resistance $ R_3 $. The parameter  
$\beta = 1 + (R_3/R_1)$. 
The noises $\eta_1(\tau)$ and $\eta_2(\tau)$  are zero-mean $\delta$--correlated Gaussian white noises,  i.e.  $\langle \eta_i(\tau) \eta_j(\tau')\rangle = \delta_{ij} \delta(\tau - \tau')$  for $i, j \in \{1, 2\}$ and are linear
combinations of  noises $\xi_i(t), i = 1, 2, 3$.  The dimensionless noise
strength is $D = 4 e k_B T / \hbar \bar{I_c}$, where $k_B$ is the Boltzmann constant. 

Note that the dimensionless form (\ref{fi}) is not symmetrical with respect to the change 
$R_1\leftrightarrow R_2$. It is because of the definition of the dimensionless time (\ref{time}) which is extracted  from the first equation of the set (\ref{Kir1}) and, in consequence, the asymmetry of $V_0$ in equation (\ref{voltage}) with respect to  $R_1$ and $R_2$ .


\section{System driven by biharmonic current}

The dynamic behavior of  two Josephson junctions described by equations (\ref{fi}) 
is analogous to the overdamped dynamics  of two interacting Brownian  particles moving in spatially 
symmetric periodic structures and driven by external time-dependent forces. However, it is not a potential system. If we think of 
$\phi_1$ and $\phi_2$ as position coordinates of two particles, then  $\dot \phi_1$ and 
$\dot \phi_2$ correspond to the velocity of the first  and second particles, respectively. 
In this mechanical analogue, junction voltage is  represented by particle velocity and time-dependent current $I_i(\tau)$
is  represented by external force.  We can use this analogue to visualize junction dynamics.
The main transport characteristics of such mechanical systems are: the long-time  averaged velocity 
 $ v_1 =\langle \dot{\phi_1} \rangle$ of the first particle and the long-time  averaged velocity   $v_2=\langle \dot
\phi_2 \rangle $ of the second particle. The question is under what circumstances one can induce    transport characterized by directed motion of both particles in the stationary regime, i.e. when 
$v_1 \neq 0$ and $v_2 \neq 0$. 
In terms of the Josephson junction system (\ref{fi}) it corresponds to the dimensionless 
long-time averaged dc voltages $ v_1 =\langle \dot{\phi_1} \rangle$ and $v_2=\langle \dot
\phi_2 \rangle $ across the first and second junction,  respectively.

 \begin{figure}[bf]
	\centering
	\includegraphics[width=0.9\linewidth]{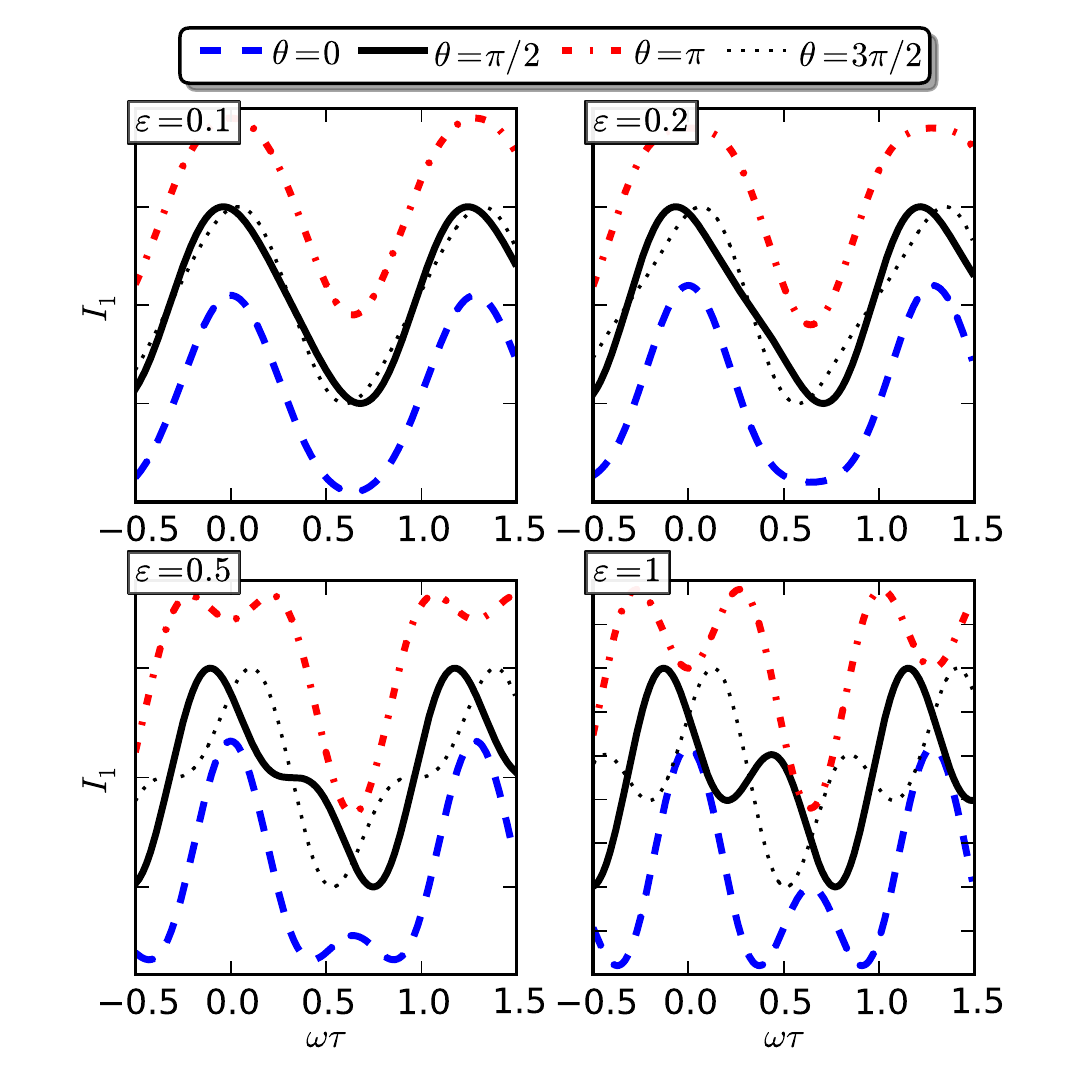}
	\caption{(color online). Illustration of the  biharmonic current (\ref{I_1}) for $ a_1 = 1 $ 
	and selected values of other parameters. Only symmetric ($\theta = 0, \pi$) and antisymmetric  ($\theta = \pi/2, 3\pi/2$) forms of the  current  are 	
	displayed for specific values of the relative phase $\theta$. 
	For clarity, the curves are offset vertically from one another to facilitate comparison.}
	\label{fig2}
\end{figure}

Transport in symmetric periodic systems like (\ref{fi}) can be trivially induced by biased external drivings $I_1(\tau)$ or/and   $I_2(\tau)$. A non-trivial case is when  zero-mean (non-biased)   driving can induce transport. An example of  zero-mean driving   is the simplest harmonic signal  $I(\tau)=A \cos(\omega \tau)$.   From the symmetry property of  equations  (\ref{fi}) it follows that such a  current source is unable to induce non-zero averaged voltages, for details see \cite{flachepl,MacLuc2010,Denisov2010,sym,niurka,renzoni}.   However, when the system is driven by  external currents having two harmonics, i.e. when \cite{schneider} 
\begin{eqnarray}
  \label{I(t)}
	I_i(\tau) = A_i \left[ \cos (\omega_i \tau) + \varepsilon_i \cos ( \Omega_i \tau + \theta_i) \right], 
	\quad i=1, 2, 
\end{eqnarray}
then transport can be induced within tailored parameter regimes.

In the paper, we will study a particular case when the  biharmonic ac current (\ref{I(t)}) is applied  only to one (say,  the first) junction, and has the form  
\begin{eqnarray}
  \label{I_1}
	I_1(\tau)  = a_1 \left[ \cos (\omega \tau) + \varepsilon_1 \cos ( 2 \omega \tau + \theta) \right], 
	\quad I_2(\tau)=0. 
\end{eqnarray}

It is known that without coupling to the second junction (i.e. when $\alpha =0$), the averaged dc voltage $v_2$ across 
the second junction is zero while the dc voltage $v_1$ across the first junction is non-zero in 
some regions of the parameters space. We can formulate several fundamental questions: (a) What 
is the influence of  coupling $\alpha$ on the voltage $v_1$? (b) Can  the voltage across 
the second junction be induced by coupling (\ref{alfa}) and how it depends on its strength?  
 (c) Can the voltages reversal be obtained by changing the control parameters? (d) Can the 
 voltages $v_1$ and $v_2$ have opposite signs?  
Properties of the ac current  $I_1(\tau)$  determine whether  
the voltages $v_1$ and $v_2$ display the above mentioned features.  
One  can distinguish two special cases   of the  two-harmonics ac current $I_1(\tau)$:  \\
(i)  The first case  is when there is  such  $\tau_0$  that $I_1(\tau_0 +\tau) = I_1(\tau_0 - \tau)$.  It  means that the driving is symmetric or invariant under the time-inversion transformation. 
It is the case when $\theta =\{0, \pi\}$, see dashed and dotted-dashed  lines in  
figure  \ref{fig2}. \\ 
(ii)  The second case is when  there is  such $\tau_1$ that $I_1(\tau_1 +\tau) = - I_1(\tau_1 - \tau)$. This is the case of the  antisymmetric driving realized for the phases 
$\theta =\{\pi/2, 3\pi/2\}$,  see solid and dotted   lines in figure  \ref{fig2}.

The analysis of a  similar system biased by a  dc current  and driven by an  unbiased  
harmonic ac signal has  been recently presented in Ref.  \cite{JanLuc2011}. 
The main difference between these two set-ups is a  constant dc current which
in turn may lead to the phenomenon of the negative mobility (resistance) \cite{MacLuc2007}. Here,
there is no such constant current applied to the system.  
However, by applying asymmetric ac signals we hope to be able to control the behavior of the first 
and second junctions. We would like to convince reader that  transport across the junction 
can be qualitatively controlled just by adjusting the \emph{shape} of the external signal 
applied to \emph{one junction only}.

\section{Voltage response to biharmonic current}\label{sec:control}

To reduce a number of parameters of the model, we consider a special case of two {\it identical} junctions. In such a case 
$ I_{c1} = I_{c2} = 1, R_1 = R_2, \alpha \beta = 1 $ and  
 equations (\ref{fi}) take the form 
\begin{eqnarray}
  \label{fi10}
    \dot{\phi}_1 = I_1(\tau)-  \sin \phi_1 
      - \alpha \sin \phi_2 + \sqrt{D}\; \xi'(\tau),  \nonumber\\
    \dot{\phi}_2 =  -  \sin \phi_2 
    	       + \alpha [ I_1(\tau) - \sin \phi_1] + \sqrt{D} \;\xi''(\tau), 
\end{eqnarray}
where the driving  current $I_1(\tau)= I_1(\tau; a_1,  \varepsilon_1, \theta)$ is given by the relation (\ref{I_1}). 

Now, we make  four general conclusions about the dc voltages as a  consequence of the symmetry properites of the ac driving current   $I_1(\tau)$:  \\
(A) -- Let us consider the voltages 
$v_1=v_1(\varepsilon_1)$ and $v_2=v_2(\varepsilon_1)$ as  functions of the 
amplitude $\varepsilon_1$ of the second harmonic. If we make  the transformation 
$\varepsilon_1 \to -\varepsilon_1$  to equations (\ref{fi10}), then it follows that 
\begin{eqnarray}
  \label{symm}
	v_1(- \varepsilon_1)= - v_1(\varepsilon_1), \quad  v_2(-\varepsilon_1)=- v_2(\varepsilon_1). 
\end{eqnarray}
These relations yield  $v_1(0)= - v_1(0),  v_2(0)=- v_2(0)$ and we conclude that 
$v_1(0)= 0$ and $v_2(0)=0$ when the second harmonic is zero, i.e. when   $\varepsilon_1 =0$. \\
(B) -- The sign of voltages $v_1=v_1(\theta)$ and $v_2=v_2(\theta)$ can be controlled by the phase $\theta$.  Indeed, if 
one changes the phase  $\theta \to \theta \pm \pi$ then $\varepsilon_1 \cos ( 2 \omega \tau + \theta \pm \pi) \to 
-\varepsilon_1 \cos ( 2 \omega \tau + \theta)$ and from the relations (\ref{symm}) one gets 
%
\begin{eqnarray}
  \label{symm2}
	v_1(\theta \pm \pi)= - v_1(\theta), \quad  v_2(\theta \pm \pi)=- v_2(\theta). 
\end{eqnarray}
(C) -- Because the driving $I_1(\tau)$ is the same for $\theta \to \pi -\theta$ and for 
$\theta \to \pi +\theta$, the symmetry relations    
\begin{eqnarray}
  \label{symm3}
	v_1(\pi - \theta)=  v_1(\pi + \theta), \quad  v_2(\pi - \theta)= v_2(\pi + \theta) 
\end{eqnarray}
hold for any value of the  phase $\theta$. \\
(D) -- For the antisymmetric ac current, equations (\ref{fi10}) are invariant under the time reversal transformation $t\to -t$. As a consequence \cite{flachepl,Denisov2010,niurka}
\begin{eqnarray}
  \label{symm4}
	v_1(\pi/2)=  v_1(3\pi/2) = 0, \quad  v_2(\pi/2)= v_2(3\pi/2) =0.  
\end{eqnarray}

The above set  of equations (\ref{fi10}) cannot be handled by standard analytical methods used in ordinary 
differential equations.  Therefore 
 we have carried out 
comprehensive numerical simulations. We have employed Stochastic 
Runge--Kutta algorithm of the $ 2^{nd} $ order with the time step of 
$[10^{-3} \div 10^{-4}](2 \pi / \omega)$. We have chosen initial phases 
$ \phi_1(0)$ and $ \phi_2(0)$ 
equally distributed over one period $[0, 2 \pi]$. Averaging was performed 
over $10^3 - 10^6$ 
different realizations and over one period of the external driving 
 $2 \pi / \omega$. 
All numerical calculations have been 
performed using CUDA environment on desktop computing processor NVIDIA 
GeForce GTX 285. This gave us a possibility to speed the numerical 
calculations up to few hundreds times more than on typical modern 
CPUs \cite{cuda}.

We begin the analysis of transport properties of the system (\ref{fi10}) by some general 
comments about its long-time behavior. 
In the  long time limit,  the  averaged voltages $\langle {\dot \phi_i(\tau)} \rangle$  can be presented 
in the form of a series of all possible harmonics, namely,  
\begin{eqnarray}
\label{asym}
\lim_{\tau\to\infty} \langle {\dot \phi_i(\tau)} \rangle =  v_i + \sum_{n=1}^{\infty} v_i(n \omega \tau), \quad  i = 1, 2, 
\end{eqnarray}
where $v_i$ is a  dc (time-independent) component and $ v_i(n \omega t)$ are time-periodic functions of zero average over a basic period.  

When both  $a_1$  and $\varepsilon_1$ are small, the  dc components $v_1$ and $v_2$ of averaged voltages are zero in the long time limit. It can be inferred from the structure of  equations (\ref{fi10}): at least one of the amplitudes  of the driving current $I_1(\tau)$ should be greater than the amplitude $I_{c1}=1$ of the Josephson suppercurrent.   Also  for high frequency $\omega$, the averaged dc voltages are zero: very fast positive and negative changes of the driving current cannot induce the dc voltage and only multi-harmonic components of the voltages can survive. For smaller values of the frequency and higher values of amplitude, one can observe  a stripe-like structure of non-zero values of both  voltages $v_1$ and $v_2$. 

We can identify four remarkable and distinct  parameter regimes  where: \\
(I) $v_1 > 0$ and $v_2 >0$,  \\
 (II) $v_1 < 0$ and $ v_2 <0$,   \\
(III) $v_1 < 0$ and $ v_2 > 0$, \\
 (IV) $v_1 > 0$ and $ v_2 < 0$. \\
   The regimes (I) and (II) dominate in the parameter space.  If the regime (I) is detected  then  the regime (II) can be  obtained from the relations (\ref{symm2}) by changing the relative phase $\theta$ of the biharmonic current $I_1(\tau)$.  Likewise, if the regime (III) is found 
then  the regime (IV) can be  determined from the relations (\ref{symm2}).

\begin{figure}[htbf]
	\centering
	\includegraphics[width=0.49\linewidth]{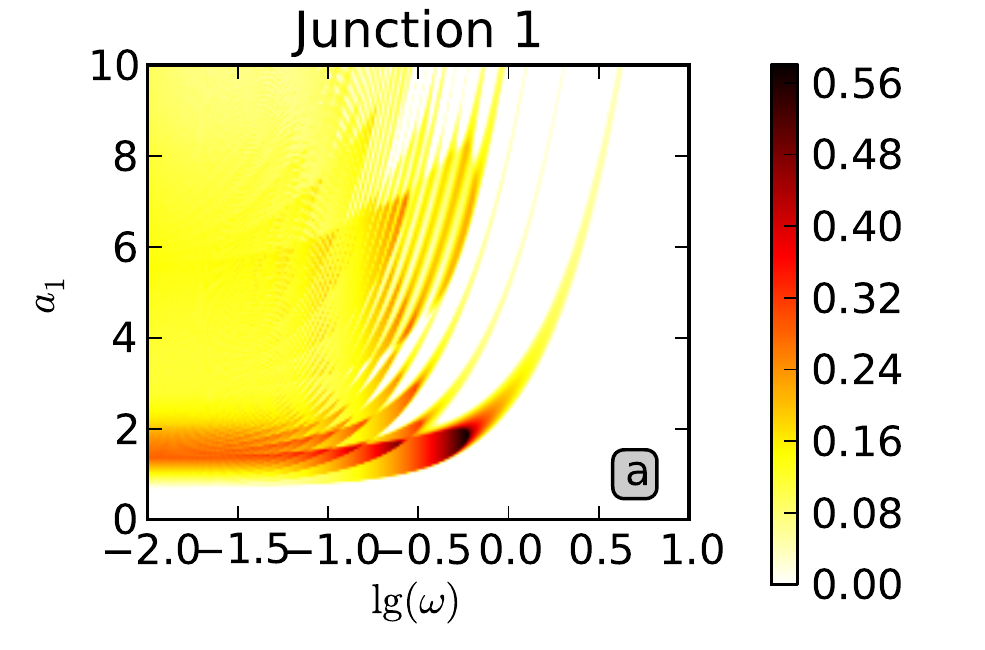}
	\includegraphics[width=0.49\linewidth]{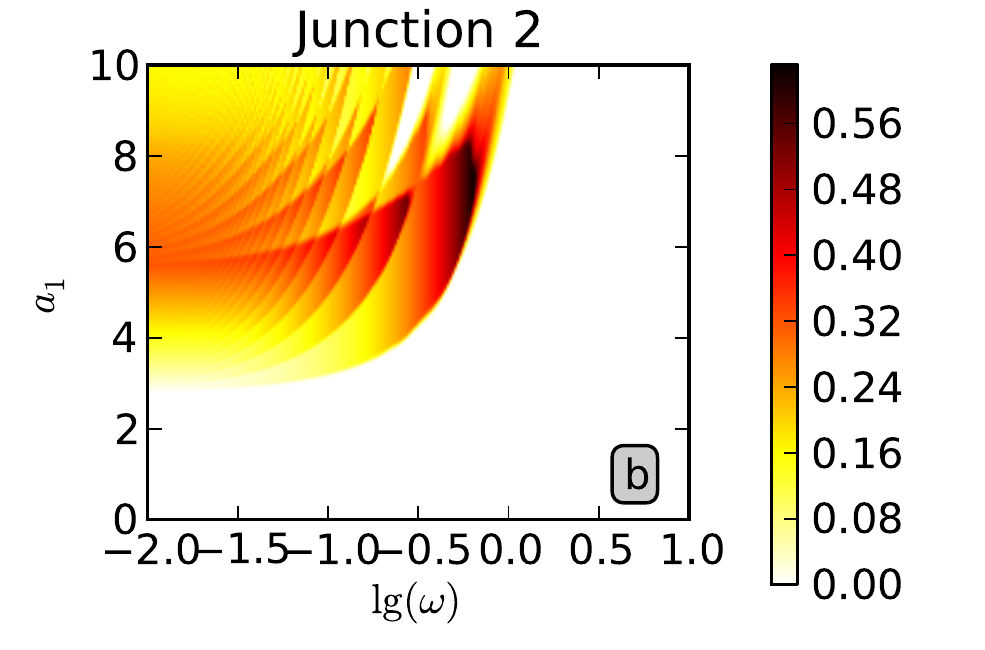}
	\caption{(color online). Transport properties of the  
  system of two Josephson junctions driven by the ac current acting on the first junction. 
   In  panels a and b averaged dc voltages $ v_1 $ and 
  $ v_2 $ across the first and second junction are shown for the coupling constant $\alpha = 0.25$, temperature 
	$D=0.001$, relative amplitude of the second harmonic $\varepsilon_1 = 0.5$ and the relative phase $\theta = 0$.}
	\label{fig3}
\end{figure}

\begin{figure}[htbf]
	\centering
	\includegraphics[width=0.49\linewidth]{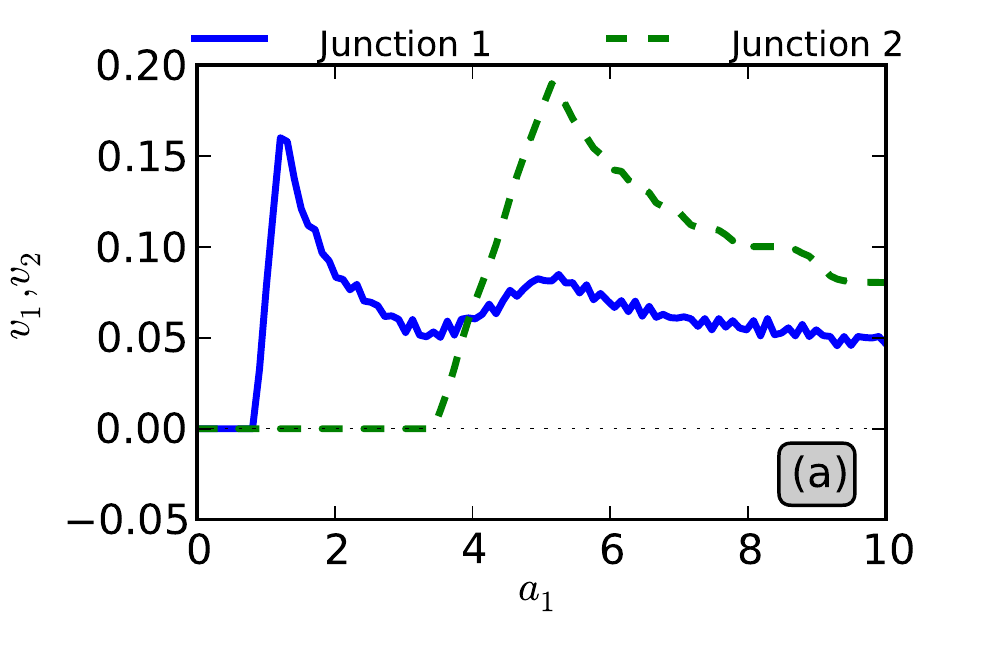}
	\includegraphics[width=0.49\linewidth]{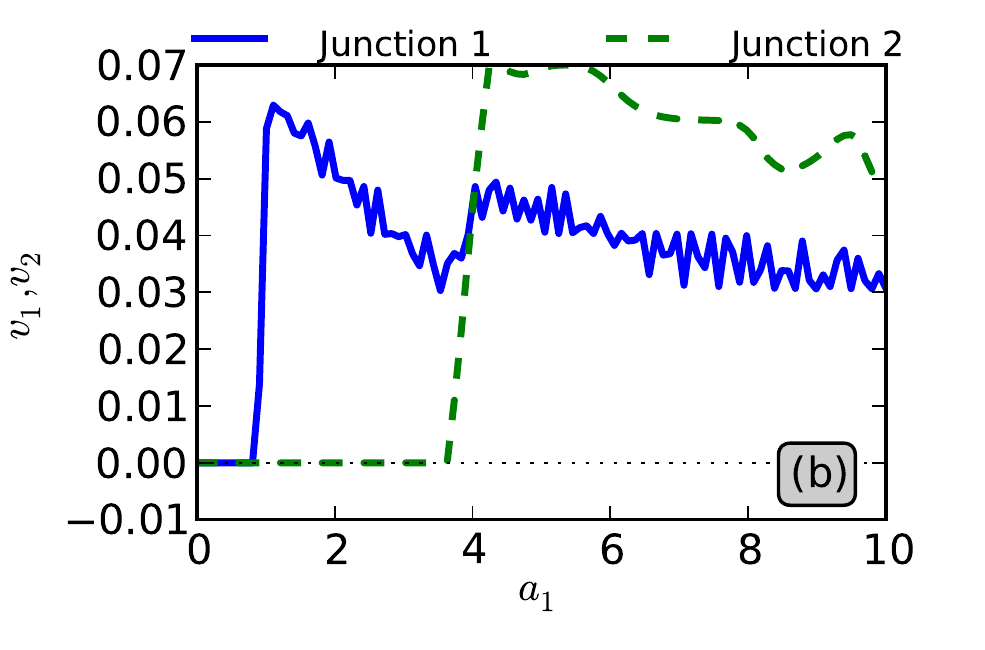}
	\includegraphics[width=0.49\linewidth]{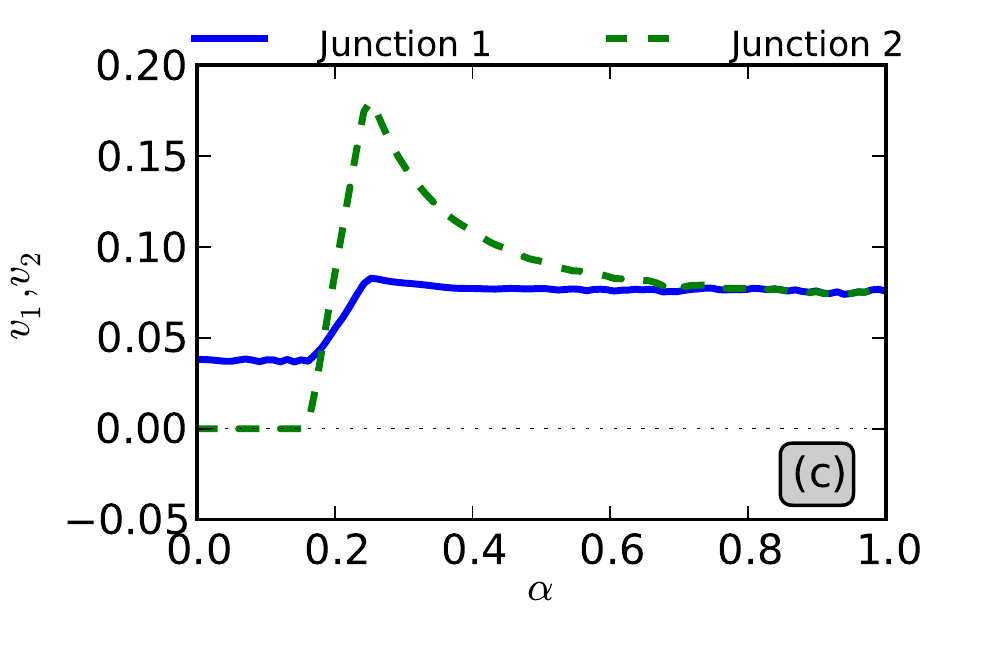}
	\includegraphics[width=0.49\linewidth]{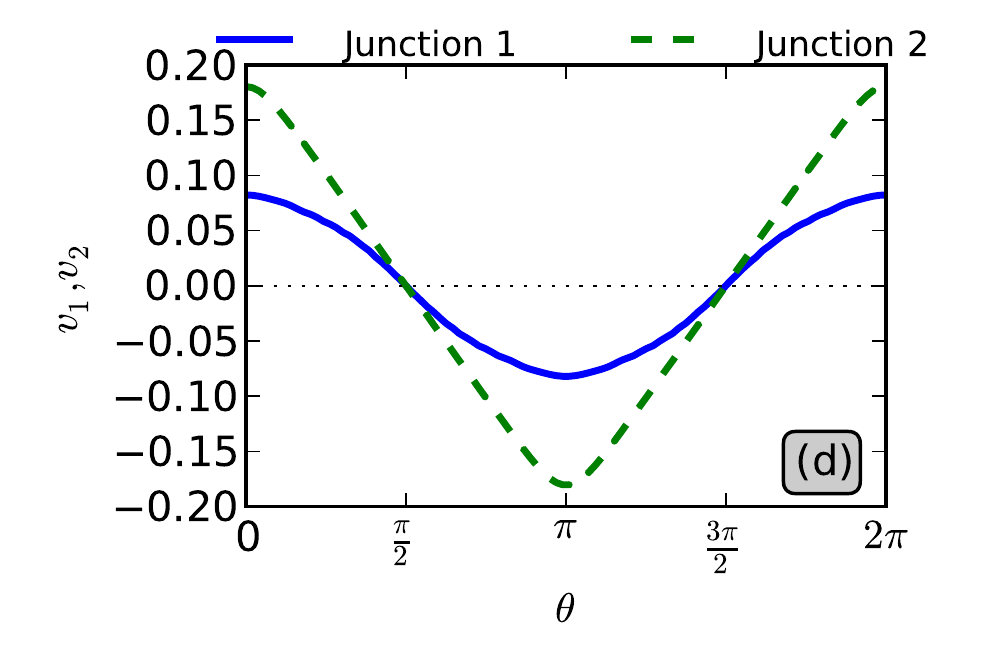}
\caption{(color online). The stationary averaged dc voltages $v_1$ and $v_2$ across 
  the first (blue solid line) and second (green dashed line) junction, 
  respectively. The parameters are: the dimensionless temperature $D=0.01$, the frequency $\omega = 0.01$ and strength  of the second harmonic $\varepsilon_1 = 0.2$.  The dependence on the external ac current amplitude $a_1$ is 
  presented in panels: (a) for the relative  phase $\theta = 0$ and (b) for $\theta =\pi / 3$, and fixed coupling constant $\alpha = 0.25$.  
  In  panel (c) the dependence of voltages on the coupling constant $\alpha$ is depicted 
  for  fixed amplitude $ a_1 = 5.15 $ and phase $\theta = 0$. 
  In panel (d) the role of the driving current  symmetry controlled by the phase $\theta$ is shown for $a=5.15$ and $\alpha =0.25$.
  }
	\label{fig4}
\end{figure}

\subsection{Dominated regime:  the same sign of dc voltages $v_1$ and $v_2$}

In figure \ref{fig3} we present  the regime (I) in the parameter plane $\{a_1, \omega\}$  which illustrate how voltages $v_1$ and $v_2$ depend on the  amplitude $a_1$   and frequency $\omega$ of the biharmonic current $I_1(\tau)$ when other parameters are fixed.  
In this regime, both voltages always take non-negative values. 

The dependence of voltages $v_1$ and $v_2$ on the current amplitude $a_1$ is depicted  in  panels (a) and (b) of figure \ref{fig4}.  
We observe that for smaller  amplitudes of external ac current the transport is  activated 
only on the first  junction while the voltage across the second junction is zero. 
We can identify non-zero  response of the first junction 
at the amplitude around $ a_1 = 1 $. The second junction 
awakes for larger values of  amplitudes (the threshold is a little bit less than $ a_1 = 4 $). 
One can note  non--monotonic behavior of all presented curves. 
After initial burst the amplitude of the voltage of the first junction decreases, 
revealing next enhancement together with the appearance of the non--zero voltage excited 
on the second junction (green dashed lines). For larger  amplitude of the ac driving, 
voltages of both junctions decrease. 
 In panels (a) and (b) of this figure, the phase $\theta$ of the second harmonic of the external 
signal is  different: $ \theta = 0$ in panel (a) and  $\theta= \pi/3$ in panel (b). The inspection of the results reveals that the change of the phase  from $ \theta = 0$ to $\theta= \pi/3$ 
reduces the dc voltages more than twice. 

 In panel (c) of figure 4, the role of  coupling is illustrated: for small values of the coupling constant $\alpha$ the voltage $v_2$ is zero. There is a threshold value $\alpha_c$ of coupling 
 above which the voltage across the second junction can be generated. 
In the regime presented in figure 4, the threshold value $\alpha_c \approx 0.152$.  
 From equation (\ref{alfa}) it follows that the external shunt resistance  $R_3$ can be a good control parameter which decides whether $v_2$  takes non-zero or zero values. For fixed remaining  system parameters,  one can induce non-zero voltage across the second junction  by decreasing the resistance $R_3$ to values $R_3 < R_c = (\alpha_c^{-1} -1) R_2$. For $R_3 > R_c$  the voltage $v_2$ diminishes.

 In panel (d) of figure 4, the influence of the relative phase $\theta$ of two harmonic signals 
 on the dc voltages is depicted. It is a numerical evidence for our consideration on  symmetry relations (\ref{symm})-(\ref{symm4}). We can deduce two conclusions: \\
 (i) The maximal absolute values of voltages $v_1$ and $v_2$ are generated by the {\it symmetric} ac current, i.e. when $\theta = 0$ or $\pi$; \\
 (ii)  Both  dc voltages $v_1$ and $v_2$ are zero when  the  ac current
is  {\it antisymmetric}, i.e. when $\theta = \pi/2$ or $3\pi/2$.\\
 From Refs \cite{flachepl,niurka} it follows that for small amplitudes of both harmonics of the driving current (\ref{I_1}) the dc voltages depend on the phase as 
$v_i \propto \cos \theta \; (i=1, 2)$.  Such a behavior is universal and does not depend on details of systems or models. E.g. it  has been observed in transport of flux quanta in 
superconducting films, see panel (b) in figure 1 of Ref.  \cite{nori}.

As follows from equations (\ref{symm2}), the regime (II)  can be obtained from the regime (I) 
 by change of  the phase  $\theta \to \theta \pm \pi$. Then the corresponding figures can be obtained from figures 3 and 4 by the reflection with respect to the horizontal axis.    Therefore we do not discuss this regime and we do not show corresponding figures.


\subsection{Regime of the opposite sign of dc voltages $v_1$ and $v_2$} 

Now, we address the question of whether we  can  identify parameter regimes (III) and (IV)  where the dc voltages $v_1$ and $v_2$ take  opposite sign. An illustrative example of such a regime is depicted in figure 5.  We can identify a fine stripe-like structure of regions of non-zero dc voltages
sparsely distributed in the parameter space.  Outside the stripes, in large regions of parameter space, the voltages $v_1$ and $v_2$ are negligible small or zero.  

 The inspection of figures 5 suggests to perform a  more accurate search for the horizontal section $a_1=const.$ and the vertical section $\omega = const.$.  
In   panels (a) and (b) of figure \ref{fig6} we present results for the section $a_1=3$. 
One can note the unique feature of the  
occurrence of windows of the  frequency $\omega$ where the voltage $v_2$  takes the 
opposite sign to the voltage $v_1$. The  voltages  as a function of 
$\omega$ display the  non-monotonic dependence exhibiting maxima and minima. We can 
detect the interval where absolute values of voltages operate synchronously: simultaneous 
increase and decrease both of them. 
For higher temperature $ D = 10^{-2} $ and within the presented range of the base 
frequency $\omega $ the voltages across  both junctions 
alternate showing very interesting and rare curves of opposite 
voltage reversals \cite{kosluc2001}. 
However, if we consider ten times lower temperature, i.e. $ D = 10^{-3} $, then only
one window of the states working contrariwise survives. It means that other windows 
are created  by thermal equilibrium fluctuations.

Two panels (c) and (d) in   figure \ref{fig6} illustrate the section $\omega = 0.1139$ and the section  $\omega = 0.1288$ to show the influence of the amplitude $ a_1 $ 
of the ac driving. The  voltages  as a function of 
$a_1$ are non-monotonic and local  maxima and minima can be detected.

\begin{figure}[htbf]
	\centering
	\includegraphics[width=0.48\linewidth]{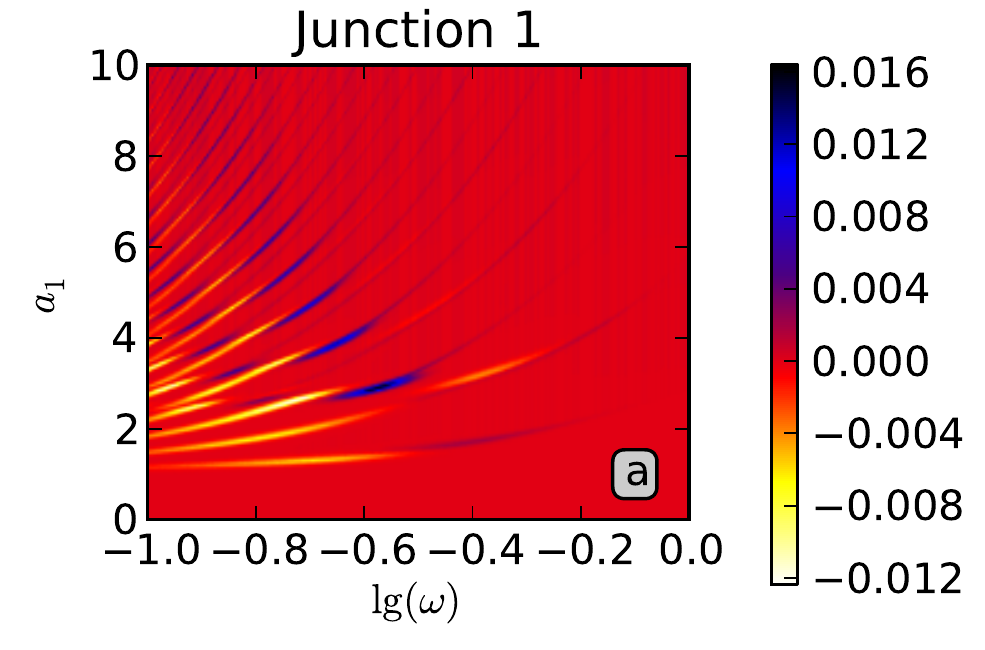}
	\includegraphics[width=0.48\linewidth]{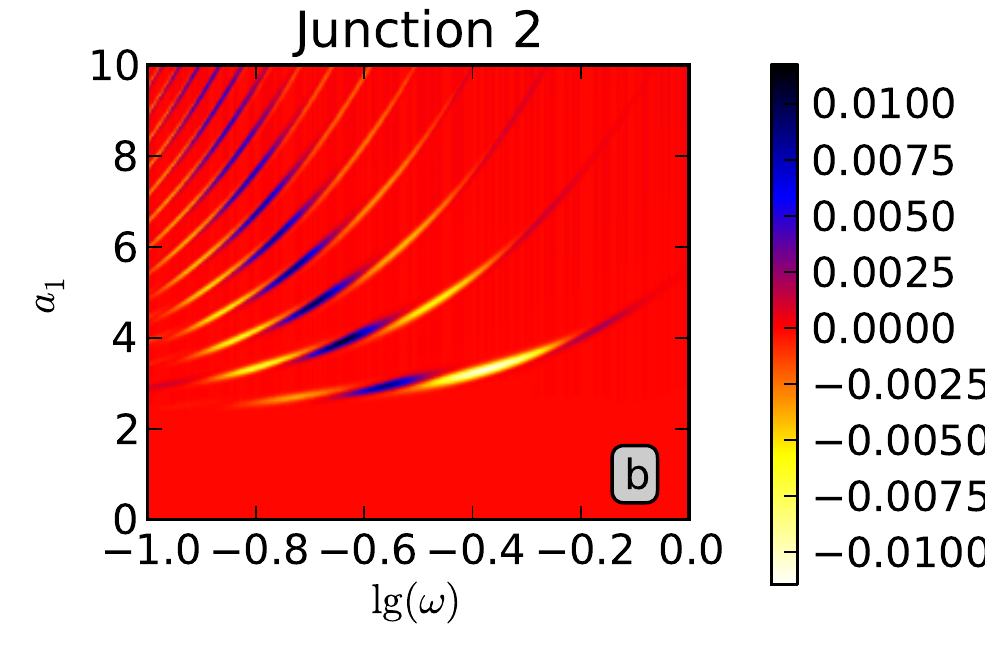}
\caption{(color online). Transport properties of the  
  system of two Josephson junctions driven by the ac current acting on the first junction. 
 In  panels a and b  dc voltages $ v_1 $ and 
  $ v_2 $  are shown for the parameters set  $\alpha = 0.25$, 
	$D=0.01, \varepsilon_1 = 0.1$ and $ \theta = \pi/2$ revealing the 
	most interesting response of junctions with characteristic stripe-like structure of regions of non-zero dc voltages. }
	\label{fig5}
\end{figure}
\begin{figure}[htbf]
	\centering
	\includegraphics[width=0.49\linewidth]{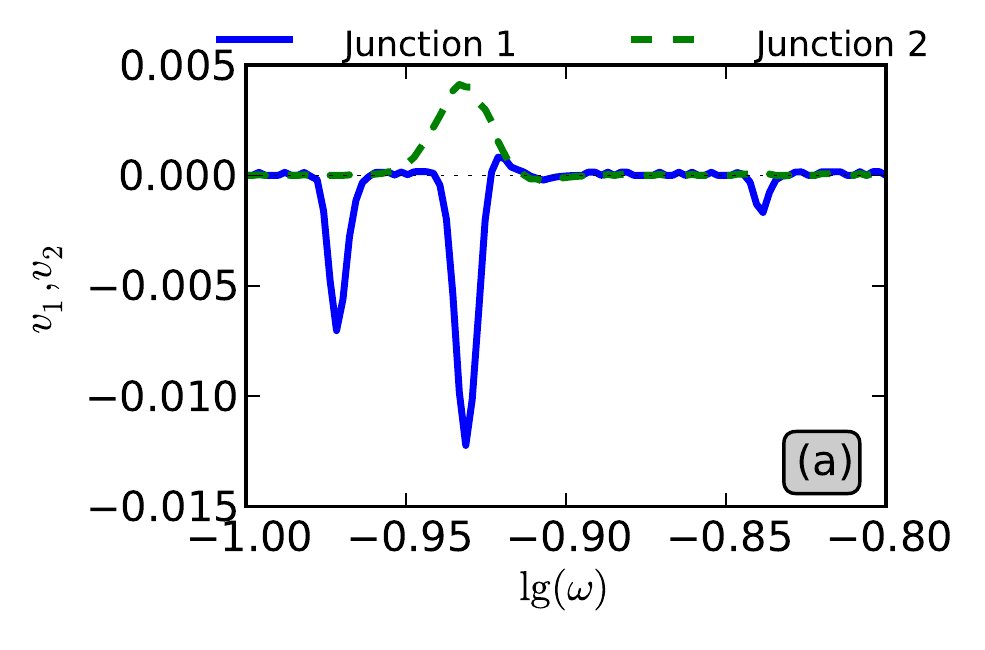}
	\includegraphics[width=0.49\linewidth]{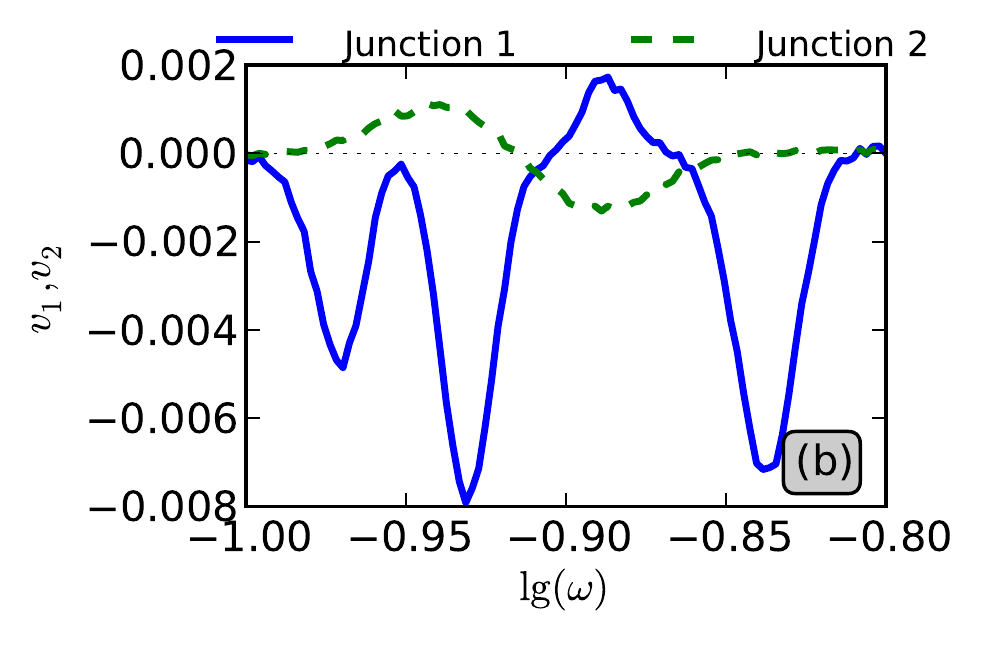}\\	
		\includegraphics[width=0.49\linewidth]{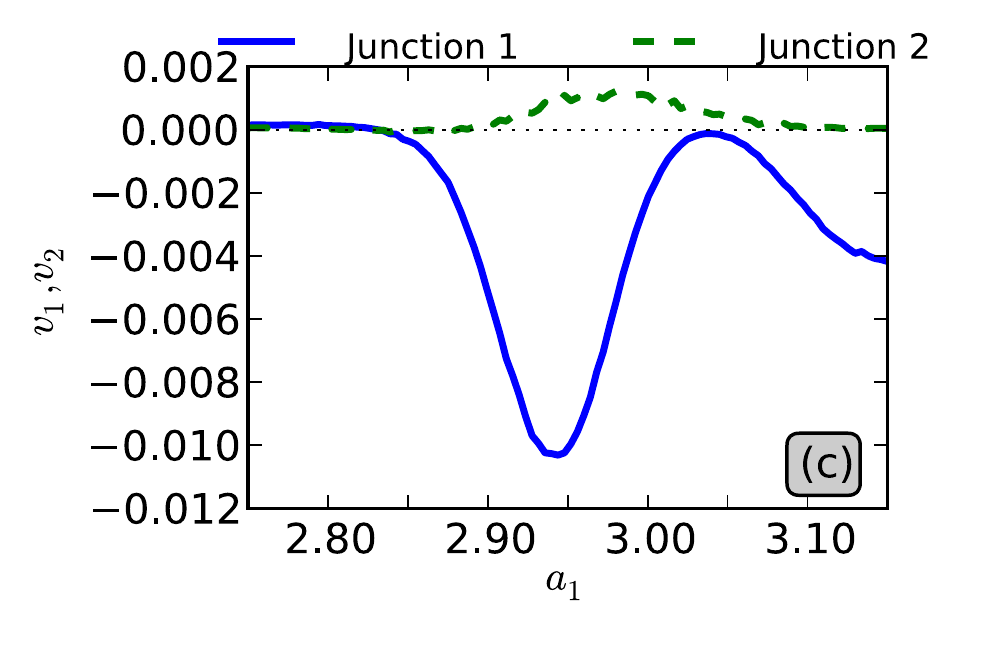}
	\includegraphics[width=0.49\linewidth]{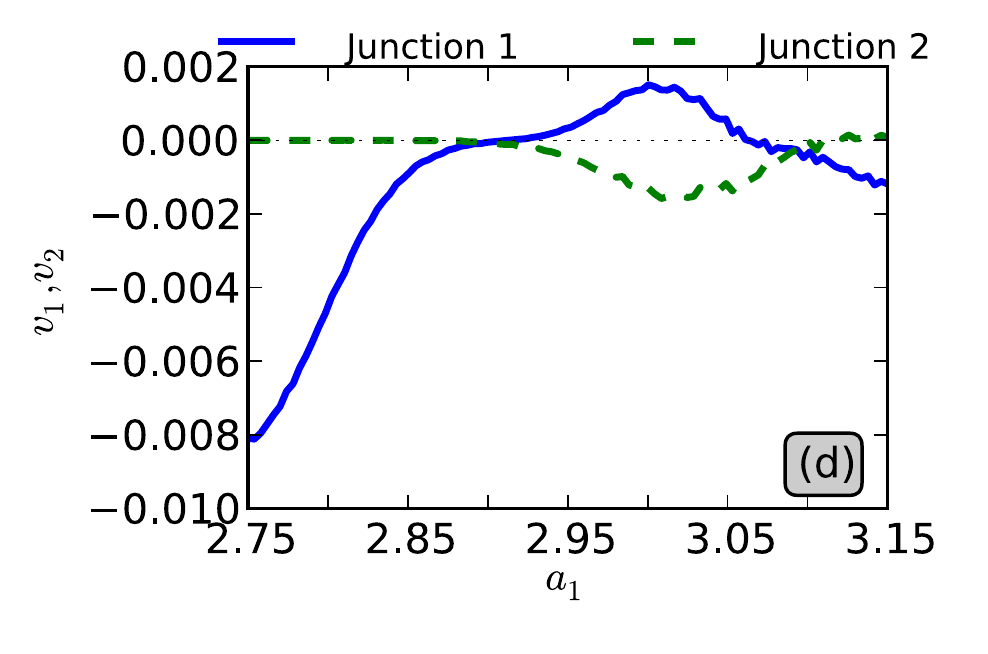}		
\caption{(color online). The stationary averaged dc voltages $v_1$ and $v_2$ across the first 
	(blue solid line) and second (green dashed line) junction, respectively. The dependence 
	on the frequency $\omega$ of the ac driving is presented in  panels (a) and (b) at a fixed 
	amplitude $a_1 = 3.0$.
	The temperature read $D = 10^{-3}$ on panel (a) and $10^{-2}$ on panel (b).
	In two lower panels (c) and (d) the dependence on 
	the external ac current amplitude $ a_1 $  is
	presented at fixed frequency $ \omega = 0.1139 $ (c) and $ \omega = 0.1288 $ (d) for  temperature $ D = 10^{-2} $.  Parameters used in all panels are: the coupling constant
	$\alpha = 0.5$, the relative amplitude of the second harmonic $\varepsilon_1 = 0.1$ 
	and the relative phase $\theta = \pi / 2$.} 
	\label{fig6}
\end{figure}

\begin{figure}[htbf]
	\centering
		\includegraphics[width=0.49\linewidth]{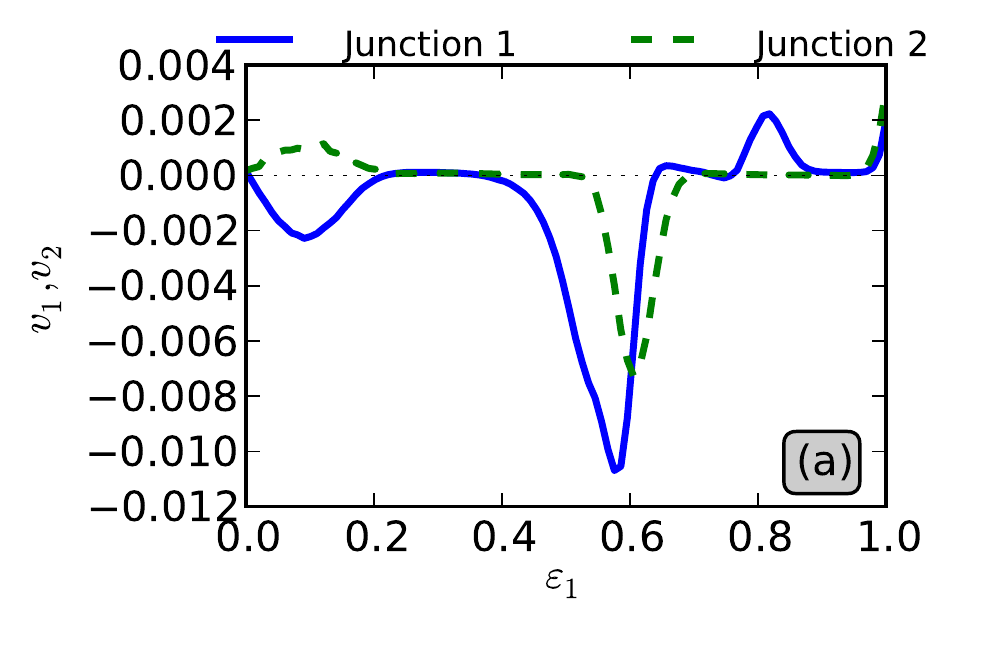}
	\includegraphics[width=0.49\linewidth]{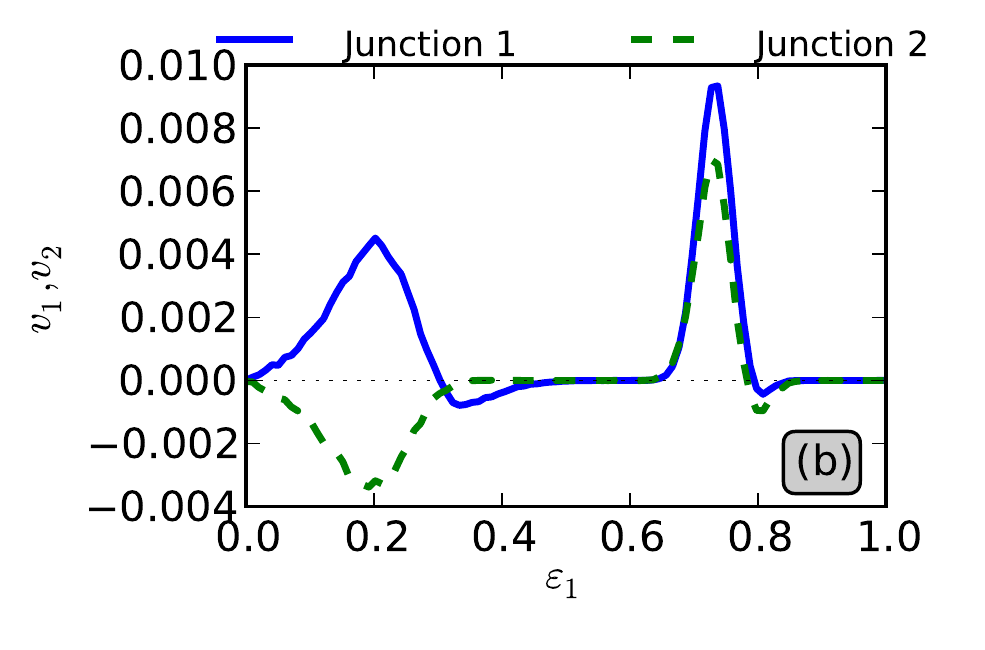}		
	\includegraphics[width=0.49\linewidth]{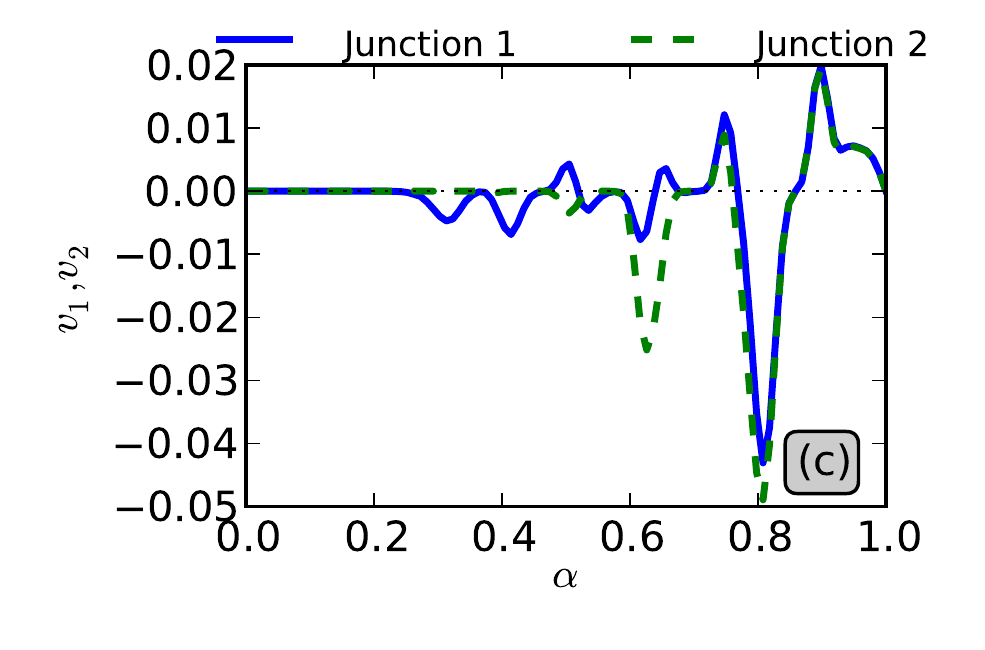}	
	\includegraphics[width=0.49\linewidth]{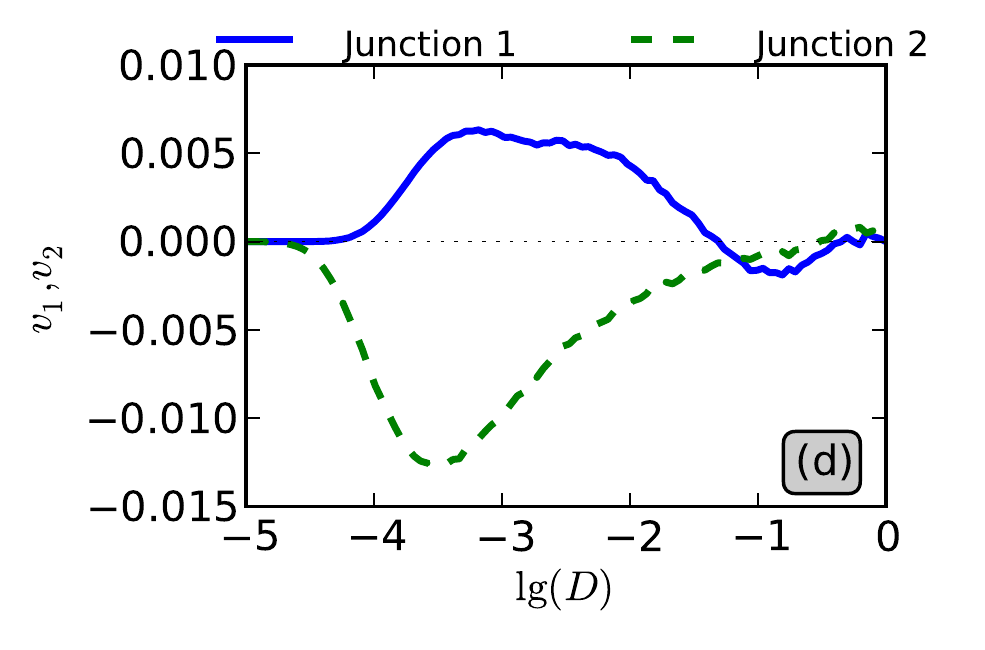}	
	\caption{(color online). The stationary averaged voltages $v_1$ and $v_2$ across 
  the first (blue solid line) and second (green dashed line) junction, 
  respectively. The dependence on the relative amplitude $ \varepsilon_1 $ of the second harmonic of the ac 
  driving is presented in panels (a) and (b) for  fixed 
  coupling $\alpha = 0.5$ and for two frequencies $ \omega = 0.1139 $ (a) and 
  $ \omega = 0.1288 $ (b). 
In  panel (c) the dependence of voltages on the coupling constant $\alpha$ is depicted 
  for  $ \varepsilon_1 = 0.2 $ and  $ \omega = 0.12884$.  
  In panel (d) the role of temperature controlled by the dimensionless parameter $D$ 
  is shown for $\varepsilon_1 = 0.2, \omega = 0.1288$ and $\alpha = 0.5$. 
      Parameters used in all panels read: relative phase 
  $\theta = \pi / 2$ and  amplitude $a_1 = 3.0$.  In  panels  (a)-(c) temperature $D = 10^{-2}$. } 
	\label{fig7}
\end{figure}

The influence of the relative amplitude  $ \varepsilon_1 $  of 
the second harmonic is depicted  in figure \ref{fig7} for two values of 
the driving frequency $ \omega = 0.1139 $ (a) and 
$ \omega = 0.1288 $ (b). In the vicinity of 
$ \varepsilon_1 = 0.1 $ we can notice previous findings where the 
voltages have the opposite sign for both junctions. However, if we increase $ \varepsilon_1 $ then  both voltages diminish and next  take negative values
following by positive signs for large $ \varepsilon_1 $, see panel (a) for details. For 
faster current  $ \omega = 0.1288 $
the situation is opposite - for larger $ \varepsilon_1 $  voltages  become positive, next negative and
end up with locked state for relatively large $ \varepsilon_1 $.
It is worth to mention that $ \varepsilon_1 = 0 $ relates to the situation where the second
harmonic is zero  resulting in a pure symmetric driving  with no possible net
transport. On the other hand, if $ \varepsilon_1 = 1 $ then both harmonics of the signal 
(\ref{I_1}) are equally strong and we are not able to say which part of it influences
the dynamics of junctions more effectively.

 In panel (c) we show the influence of interaction 
between two junctions on the voltage characteristics. The first observation is that there 
are optimal values of the coupling strength $\alpha$ for which absolute values of  
voltages are locally maximal. The second observation is that  variation  of the coupling strength 
can lead to the phenomenon of the voltage reversal \cite{kosluc2001}: the voltage changes its values from 
positive (negative) values to negative (positive) ones.     
The third observation is the same as presented in panel (c) of figure 4: there is 
 a threshold value of coupling $\alpha=\alpha_c$ (or the resistance $R_3=R_c$)  such that for 
 $\alpha < \alpha_c$ (or for $R_3 > R_c$)  the voltage $v_2$ is reduced to zero.

 In panel (d) we depict  the influence of temperature.  
 Most remarkably, transport  is solely induced by thermal noise. Indeed, in the deterministic case (i.e. for $D=0$), when no thermal fluctuations act,  the dc voltages vanish.   With increasing temperature, the voltage $v_1$  starts to increase to positive-valued local maximum. 
 Next, it decreases crossing zero and reaches a negative-valued local minimum. For further increase of temperature, the voltage $v_1$  tends to zero. 
In turn,  the voltage $v_2$ takes negative values  reaching minimal value as temperature start to increase. Next, it increases crossing zero  and reaches local maximum. Finally, it monotonically  decreases toward zero. 
 We observe that optimal temperature occurs at which the voltage $v_1$ is maximal. There is also 
another optimal temperature at which in turn the absolute value of $v_2$ is maximal. 
Moreover, one can identify the voltage reversal phenomenon: the both voltages change their sign as 
temperature is varied.

From the presented analysis it follows that  a biharmonic drive allows one to conveniently manipulate transport with high efficiency  by changing the system parameters.

\section{Summary}\label{sec:summary} 

This paper presents the detailed study of the system of two coupled noisy Josephson 
junctions which undergoes the influence of the external ac biharmonic current applied to one 
of the junctions only. The dependence of voltages across the first and second junctions upon 
ac driving exhibits a rich diversity of transport characteristics. In particular, 
it is  possible
to control both junctions to operate simultaneously with positive or negative voltages of the 
same sign or,  more interestingly, with the opposite sign. 
Moreover, the phenomenon of the voltage reversals is revealed: the voltages change the sign as one of
the parameters is varied.
 Our findings can be experimentally verified in an accessible set-up of two Josephson junctions coupled by an external resistance.  The work can open the  perspective
of a new type of  electronic elements which is
tunable between positive and negative dc voltages via external 
control parameters like the amplitudes and relative phases 
of an ac current. 
The straightforward extension of the ideas presented here would cover an increasing
number of coupled junctions to three (or more) as well as study of 
other types of driving.

\ack
The work supported in part by the grant N202 052940.


\end{document}